# Soft granular particles sheared at a controlled volume:

# Rate-dependent dynamics and the solid-fluid transition

J.-C. Tsai [a,*], M.-R. Chou [b,a], P.-C. Huang [b,a], H.-T. Fei [a], and J.-R. Huang [a]

a)  Institute of Physics, Academia Sinica, Taipei, Taiwan

b)  Dept. of Physics, Nat'l Taiwan University

*)  jctsai@phys.sinica.edu.tw

We study the responses of fluid-immersed soft hydrogel spheres that are sheared under controlled volume fractions. Slippery, deformable particles along with the density-matched interstitial fluid are sandwiched between two opposing rough cones, allowing studies for a wide range of volume fraction $\phi$ both above and below the jamming of granular suspension. We utilize sudden cessations of shearing, accompanied by refraction-matched internal imaging, to supplement the conventional flow-curve measurements. At sufficiently high volume fractions, the settling of particles after the cessations exhibits a continuous yet distinct transition over the change of shear rate. Such changes back out the qualitative difference in the state of flowing prior to the cessations: the quasi-static yielding of a tightly packed network, as opposed to the rapid sliding of particles mediated by the interstitial fluid whose dynamics depends on the driving rate. In addition, we determine the solid-fluid transition using two independent methods: the extrapolation of stress residues and the estimated yield stress from high values of $\phi$, and the settling of particles upon shear cessations as $\phi$ goes *across* the transition. We also verify the power law on values of characteristic stress with respect to the distance from jamming $\phi - \phi_c$, with an exponent close to 2. These results demonstrate a multitude of relaxation timescales behind the dynamics of soft particles, and provoke questions on how we extend existing paradigms on the flow of a densely packed system when the softness is actively involved.

## Introduction

Soft-matter systems, in many circumstances, exhibit solid-fluid duality. They can be driven to flow steadily and accommodate indefinite amount of shear strain. Meanwhile, the same systems may exhibit a solidity with a resistance to shear, namely, a yield stress. Monitoring the development of yield stress has been a common criterion for the emergence of solidity over the change of controlling parameters including, but not limited to, the volume fraction $\phi$ or temperature of suspensions [1,2]. The solidity is sometimes presented in the form of a residual stress upon the cessation of driving, and has been widely studied in glassy systems such as in experiments with synthetic clays or microgels [3-6], or in numerical studies including both hard and soft spheres [7], revealing a wealth of phenomena involving relaxations at different timescales.

Granular materials at high $\phi$ present interesting examples showing the solid-fluid duality. In the past decades, studies using shear flows have established dimensionless quantities such as inertia number and viscous number that successfully capture behaviors of both dry grains and particle-fluid mixture for a wide range of volume fractions $\phi$ up to 0.58 [8-11]. As reviewed by Guazzelli and Pouliquen [9], the current paradigm in theories has assumed Newtonian behaviors, i.e., the stress varying linearly with the shear rate $\dot{\gamma}$ (but with a $\phi$-dependent viscosity) at constant volumes, while most supporting experiments and numerical studies are stress-controlled. On the other hand, it is not uncommon that at values of $\phi$ near the solid-fluid transitions, particle suspensions can exhibit shear-thinning behaviors[1] such that the shear stress is proportional to $\dot{\gamma}^n$ with an index $n$ significantly less than 1 and in many cases close to 0.5 [12-14]. It is also worth noting that all hard-sphere theories would inevitably lose their predictive power on the flow behaviors at volume fractions beyond the "jamming point" -- the current consensus for frictionless particles is the random-close-packing (RCP) $\phi_{RCP} \approx 0.635$, despite some debates had existed with the exact meaning of RCP[15]. How we understand the yielding and flow of densely packed soft particles remains a profound challenge. Therefore, further experimental information on the rheological behaviors as well as the shear-induced structural changes across the solid-fluid transition is of vital importance.

In this paper, we start with an *Overview* of our experiments on shear flows of centimeter-sized hydrogel particles immersed in a density-matched interstitial fluid. This section provides a survey over a wide range of volume fractions with the conventional flow-curve measurements, providing hints on the *solid-fluid transition* and an intriguing *rate-dependent dynamics* at densities well above the jamming point. These behaviors are systematically analyzed in two subsequent sections, using sudden cessations to the steady shearing in combination with our internal imaging of particle movements. Other factors such as a *transition of fluid dynamics* that coincides with the solid-fluid transition, the *weakening of hydrogel under stress over long time* and its possible effects are reviewed in our *Discussion.* In the *Conclusion*, we summarize our findings and open questions that call for further understanding.

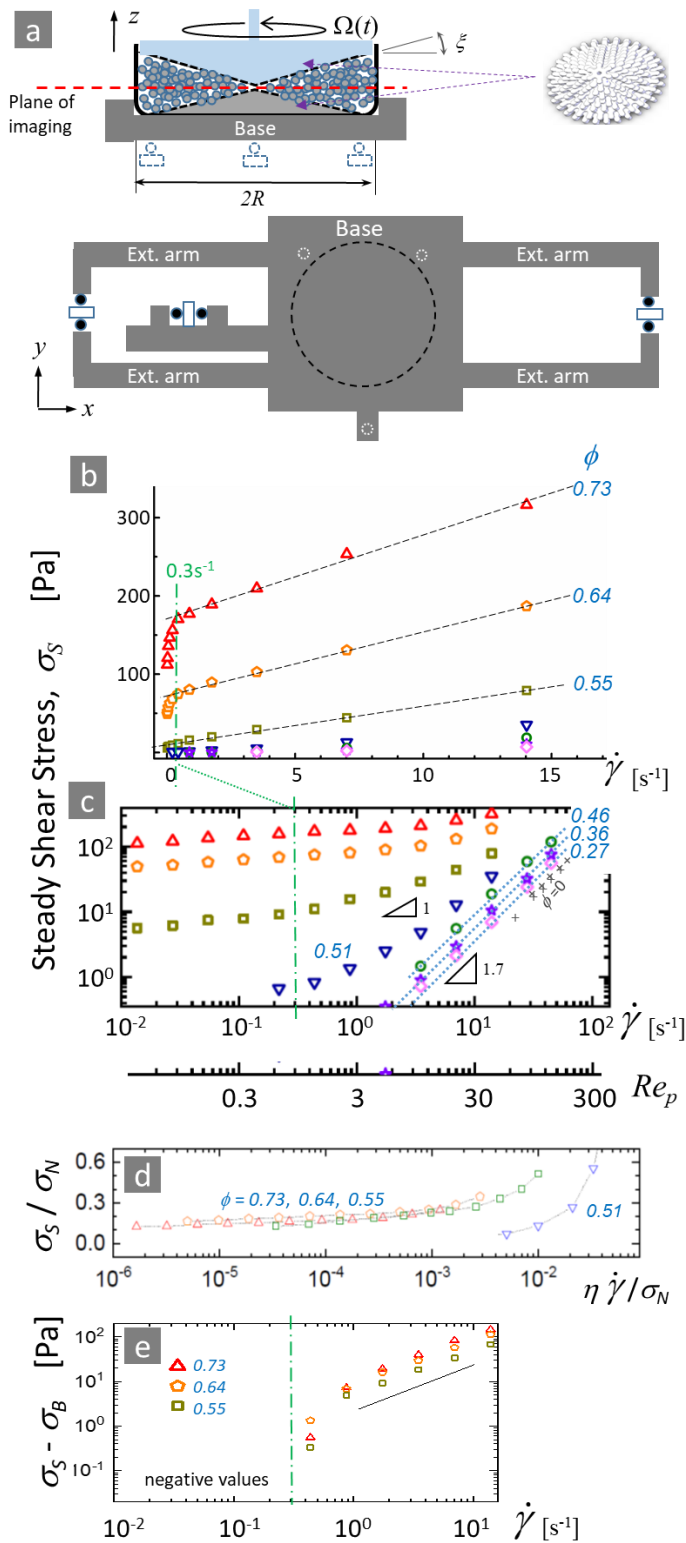

**Fig. 1 (a)** Vertical cross-section of the setup, definition of symbols, and an angled view of the cone structures. Lower part shows a top view of the base and extended arms. The base is constrained by six force sensors (empty rectangles) that are fixed on a rigid supporting system (not shown). **(b)** Time-averaged shear stress $\sigma_S$ as functions of the shear rate $\dot{\gamma}$, at different values of $\phi$. Data are time-averaged for a strain accumulation over 100 for each point. Dashed lines represents a best fit to the Bingham model $\sigma_S = \sigma_B + K\dot{\gamma}$. **(c)** Curves of $\sigma_S$ displayed in logarithmic scales. Slopes of 1 and 1.7 are shown as guides to the eyes. The corresponding particle Reynolds numbers are displayed under the logarithmic axis of $\dot{\gamma}$, as a reference. The two limiting cases with pure fluids ($\phi=0$) are displayed with crosses (x) or plus signs (+), respectively, for experiments using the standard PVP solution or with water. **(d)** Shear-to-normal stress ratios, plotted against the viscous number, for experiments at four representative values of $\phi$. **(e)** Close-up of (c) around the transition over shear rate, with their respective Bingham stress $\sigma_B$ subtracted for clarity. The solid line indicates a logarithmic slope 1 corresponding to the model. Values of $\sigma_B$ are 170Pa, 73.5Pa, and 10.8Pa, in sequence, for $\phi$=0.73, 0.64, and 0.55. The green dash-dot line marks the same shear rate 0.3s⁻¹ in (b), (c), and (e) to guide the eyes.

## Overview with steady-state experiments

Shown in Fig.1a, the soft hydrogel particles of diameter $d$=1.2cm fill the space between two opposing cones that are roughened by steps comparable to $d/2$. The cylindrical container is made of smooth glass with an inner diameter $2R$=23cm. The upper cone is set to rotate at a fixed height and with an angular speed $\Omega$ driven by a programmable stepping motor with an angular resolution of $2\pi/10^4$. Force sensors attached to the base measure the shear stress and normal stress as functions of time. The gap between the edge of rotating cone and the stationary sidewall is 2mm, and that between the two tips of the cones is 1mm, such that all particles are well contained between two cones. The *volume fraction* is determined as $\phi = N v_1 /(V_{total} - \varepsilon)$ in which $N$=O($10^3$) is the total number of particles. $V_{total}$ represents the total volume between the two cones and $v_1$ is the average volume of a single particle. The vertical steps (for generating the roughness) and the circumference of the container create certain dead volume that is not accessible, $\varepsilon \sim 0.037\, V_{total}$, for our particles with diameter $d$. Data discussed in this work are based on monodisperse hydrogel particles that are commercially available. Similar products have been characterized independently by previous works [16,17]. We have confirmed that the response of these hydrogel spheres to normal compression is well described by Hertzian model with the elastic moduli $E$ ~30kPa. In addition, we have verified that the friction coefficients of the particle-particle and particle-sidewall contacts are well below 0.01. For all experiments discussed in this paper, particles are fully immersed in the aqueous solution of 1.7% PVP-360 (polyvinylpyrrolidon) to achieve density match that prevents sedimentation. The viscosity of the PVP-360 solution is $\eta$= 8mPa-s. Simultaneously, we take internal images of particles at the mid-height of the packing, with details to be described in a later session.

### 1. Flow curves

Figure 1b shows the measured steady-state shear stress, $\sigma_S \equiv 3$ torque $/2 \pi$ R$^3$ , plotted against the effective shear rate $\dot{\gamma} \equiv \Omega/ (2 \tan \xi) = 13.7 \cdot \Omega/2\pi$ for several values of $\phi$. For $\phi$ =0.73, 0.64 and 0.55, the high-$\dot{\gamma}$ data are fitted by the Bingham model, i.e., $\sigma_S$ is linearly dependent on $\dot{\gamma}$ with a nonzero offset $\sigma_B$. However, $\sigma_S$ at low shear rates deviates significantly from the linear extrapolation of the high-shear data; measured values go significantly lower than the Bingham offset as the shear rate further decreases. Such deviation occurs around the shear rate around 0.3s$^{-1}$, for which a close-up view is shown as Fig.1(e) in logarithmic scales. This suggests a rate-dependent dynamics as the consequence a competition between the elastic, static packing and the fluid-mediated sliding. We dedicate the next section of this paper in analyzing this competition, using the settling of particle motion after sudden cessations of external shearing as the main probe.

In Fig.1c, the flow curves are displayed in logarithmic scales at a wider range of volume fractions. We also shows the *particle Reynolds number* $Re_p \equiv \rho \left(\frac{d}{2}\right)^2 \dot{\gamma} /\eta$ , which quantifies the relative importance of the *inertial force* in comparison with the *viscous stress* in a conventional suspension. In our experiments, the particles and fluids are density-matched at $\rho$ =1.2 g/c.c. and both the particle diameter $d$ and the interstitial fluid viscosity $\eta$ are fixed. Therefore $Re_p$ can be regarded simply as a dimensionless shear rate. Meanwhile, as the volume fraction decreases, particles would eventually lose direct contact and the system becomes a true suspension. While all measurements span across both the low-$Re_p$ ($<1$) and the high-$Re_p$ ($>1$) regimes, data at low $Re_p$ for low-volume fraction experiments ($\phi > 0.51$) are all below our instrumentational limit and are thus omitted from the graph. In these dilute cases, the presented data are dominated by the fluid dynamics and exhibit a simple, constant logarithmic slope with $\sigma_S \sim \dot\gamma^{1.7}$. However, this should not be interpreted as shear thickening. Based on our previous work in a similar geometry[18], we have demonstrated that these dilute flows are in the regime where fluid inertia can induce secondary vortex. The strong effect of inertia also explains why the shear stress appears insensitive to the viscosity at the limit of $\phi = 0$ in this experiment --- see the good overlap of data points for pure water ($\eta \sim 1$ mPa $\cdot$ s) and those for our standard PVP-360 solution ($\eta \sim 8$ mPa $\cdot$ s)

## 2. Stress ratios

We find that the stress ratio $\sigma_S/\sigma_N$, where $\sigma_N \equiv F_z/\pi R^2$ , shows an interesting collapse for high-volume-fraction experiments, when plotted against the *viscous number* $J \equiv \eta \dot\gamma /\sigma_N$ which weights the viscous stress over the *total pressure*. This dimensionless number has been commonly used in the studies of dense granular suspensions [9]. For $\phi$ = 0.73, 0.64, and 0.55, Fig. 1d reveals that the stress ratio $\sigma_S/\sigma_N$ is roughly rate-independent ($\approx 0.14$) in the small-$J$ regime. This observation largely agrees with a previous prediction from simulations of nearly rigid frictionless particles in the quasi-static regime [19], but ours has extended to the cases that substantial deformation of particles is expected with $\phi = 0.73 > \phi_{RCP}$. The curve of $\sigma_S/\sigma_N$ changes dramatically as $\phi$ decreases by just 0.04 (from 0.55 to 0.51), in comparison to the relatively mild change from 0.73 to 0.55.

The good collapse of data point for $\phi$ =0.73,0.64,and 0.55 seems to extend the previous finding that the stress ratio can indeed be well characterized by a single dimensionless number $J$ for multiple volume fractions in granular suspensions [9]. Nevertheless, our measurements are made for much denser packing of particles than in previous studies. Reduction of volume fraction to $\phi$ =0.51 creates a clear separation of the curve from those at higher volume fractions, in consistence with the picture that the system has become a suspension. However, a more accurate determination of the solid-fluid transition might require other indicators, which we would describe in the subsequent section with alternative measurements based on residual stress and particle movements.

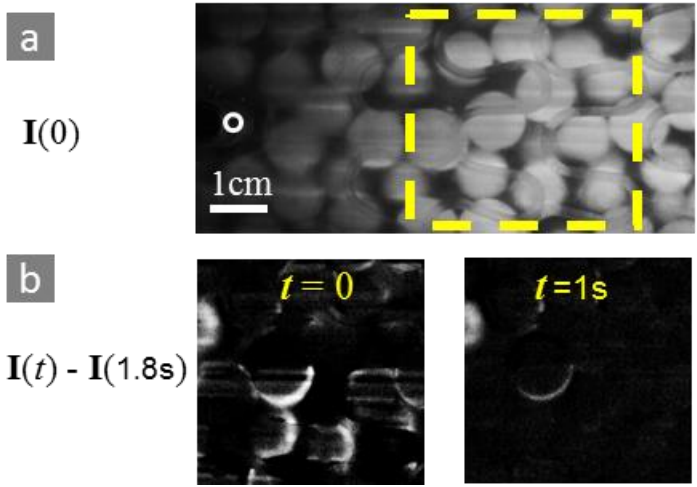

**Fig.2** Settling of particles observed along the plane of imaging. **(a)** Snapshot of an original image at a volume fraction 0.73. The empty circle on the left indicates the center of the container. **(b)** Two images subtracted by the same reference frame at $t$ = 1.8s, for the region bounded by the dashed square in (a). Timestamps are shown on the top of each image.

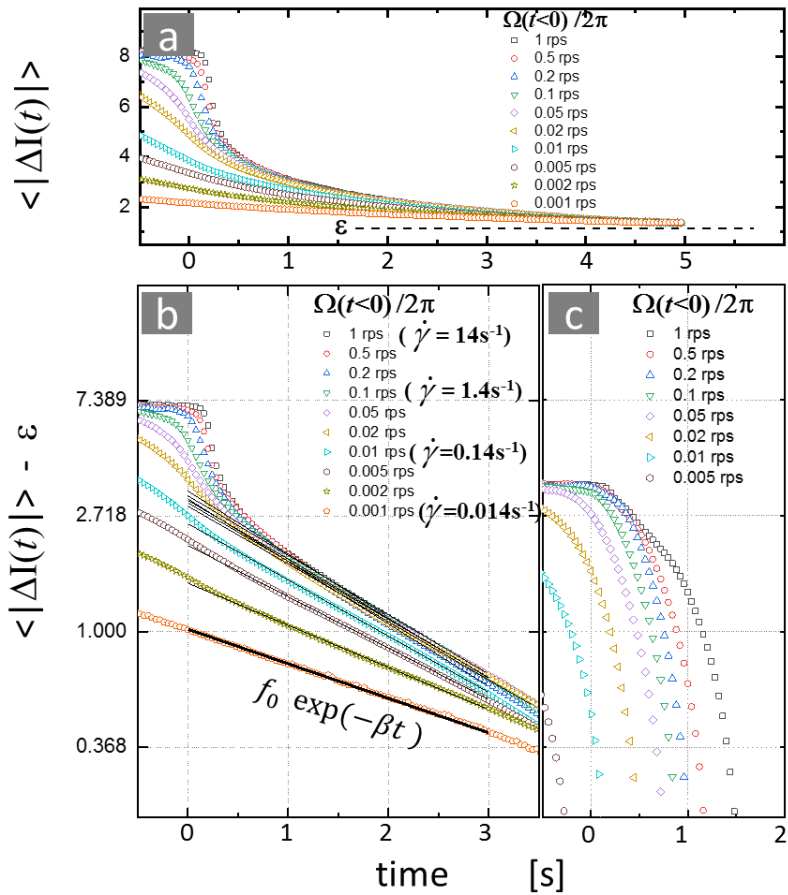

**Fig.3 (a,b)** Cumulative pixel intensity of the decay of the image difference, $\langle|\Delta\mathbf{I}(t)|\rangle$, as phase-averaged over multiple cycles of shear cessations with a reference frame set at t=5s. $\phi = 0.705$. The preshearing (t<0) are driven by a range of rotation rates $\Omega_{on}/2\pi$ spanning over three decades (in rps) -- Representative shear rates are indicated by $\dot\gamma$ in s$^{-1}$. The asymptotic of $|\Delta\mathbf{I}(t)|$ converges to a constant $\varepsilon$ that is determined empirically, corresponding to a fixed level of noise. **(c)** Same as (b), for $\phi = 0.55$

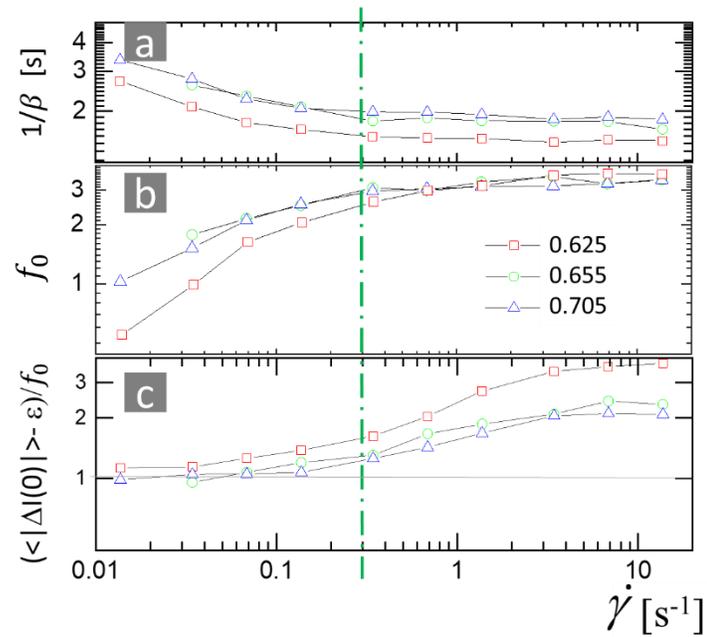

**Fig.4** Summary on the driving-rate-dependent settling of particle movements, for experiments at three volume fractions above 0.6 as indicated by the legend. **(a-b)** Values of $1/\beta$ and $f_0$ for the best fits as shown in Fig.3b. **(c)** $\langle|\Delta\mathbf{I}(0)|\rangle - \varepsilon/f_0$, featuring the degree of deviation from a simple exponential decay. All are shown as functions of the corresponding shear rate $\dot\gamma$ prior to the sudden cessation. The vertical dot-dash line marks the same 0.3s$^{-1}$ as in Fig.1(b-c) to guide the eyes.

# Rate-dependent dynamics at high volume fractions

As a way to probe the rate-dependent dynamics, we apply *sudden*[*] cessations to interrupt the steady flows, and observe the settling of particles afterwards. For this purpose, we apply fluorescence-contrast internal imaging [20,21]. Here, we stain these hydrogel particles with Nyle Blue, and illuminate them at the mid-height by a 1mm-thick laser sheet at a wavelength 635nm ---- see Fig.1a. Figure 2a shows a close-up of the bandpassed image (at 656±10nm, for the fluorescence). Figure 2b shows the subtraction by a fixed frame at $t_{stop}$+1.8s as time evolves, to illustrate the settling of particles towards the final state of quiescence. Movements of particles beyond $t_{stop}$+2s are hardly identifiable by eyes.

It's worth noting that such transient rearrangements have been predicted in numerical studies targeting the dynamics of microgels[3]. Nevertheless, our experiments present direct visual demonstrations of such effects but in the context of a centimeter-sized granular packing. As a rough indicator for these transient movements, we set the origin $t$=0 upon the cessation of shearing, subtract the time series of image $\mathbf{I}(t)$ by a reference frame set at $t$=5s, compute the mean pixel intensity of each frame and average over 40 repeated cessations. Such "phase-averaged" $<|\Delta\mathbf{I}(t)|>$ are computed for experiments at a wide range of driving rates and at different volume fractions, with imaging conditions kept identical (using the same camera with 876x876 pixels at a grayscale of 256, using the same exposure time of 8ms and a fixed illumination). Figure 3(a,b) illustrates the decay of $<|\Delta\mathbf{I}(t)|>$ in a typical case above jamming. In particular, with the noise level $\varepsilon$ removed, the semi-logarithmic plot Fig.3b reveals interesting clues on the qualitatively different dynamics behind the "slow" versus "rapid" flows, which results in distinct outcomes of the settling in response to an abrupt stopping of the boundary:

(1) *At shear rates of 0.1s$^{-1}$ and below,* the decay of particle movements is close to a simple exponential function. The curves are nearly a smooth continuation from t<0 in which the packing is driven to flow, presumably, with successive local yielding of the elastic network. The time evolution after the cessation makes little distinction on whether the boundary has stopped. However, in this "quasi-static" regime, the cumulative magnitude of relaxation ($f_0$), as well as the decay rate ($\beta$), still has a subtle dependence of on the driving rate ($\Omega_{on}/2\pi$). This subtle rate dependence and its relationship to the force balance behind the flows post delicate questions for theories.

(2) *At shear rates above 1s$^{-1}$,* the settling of particles presents a two-stage process. Upon the abrupt stopping of the boundary, $|\Delta\mathbf{I}(t)|$ shows a steep descent within the first half second. Images reveal that particles exhibit rapid movements in finding neighbors for building up a network, based on which further adjustments take a much slower pace. The follow-up adjustments, cumulatively, are in general much smaller than 1$d$ for each individual particles in reaching their final position --- see Fig.2b for two snapshots of the image $\Delta\mathbf{I}$. Interestingly, the good overlap of $<|\Delta\mathbf{I}(t)|>$ in the second stage indicates that the cumulative relaxations are, in average, insensitive to the initial shear rates. This suggests that, with sufficiently high values of $\Omega_{On}$, the combination of the rapid boundary shearing and the fluid dynamics at intermediate-to-high Reynolds numbers creates a set of fluid-like, unsupported initial configurations. It is then understandable that the outcomes of the relaxation, in average, appear insensitive to the history prior to the cessations (t<0).

---

[*] By precision optical monitoring, we have verified that our rotating boundary stops abruptly within 10ms in all cases.

In addition, for the three high-volume-fraction cases that exhibit similar behaviors, we fit the long tails as shown in Fig.3b by an exponential decay, $f_0 \exp(-\beta(t - t_{stop}))$, in which $1/\beta$ features a characteristic time of decay. The constant $f_0$ can be regarded as the magnitude of total relaxation for the second stage. Results are shown in Fig.4(a,b). In addition, with Fig.4c, the ratio $(<|\Delta\mathbf{I}(0)|>- \varepsilon)/f_0$ measures the difference of the full curve from a simple exponential decay: a large value indicates the occurrence of the aforementioned two-stage process, while unity corresponds to the case with a straight exponential. In short, Figure 4 summarizes *a gradual transition of dynamical regimes over driving rate*: the incremental yielding of elastic "solids" at the slow regime, in contrast to the more "fluidic" sliding of particles mediated by interstitial fluid flows at the fast regime. Note that such transition over driving rate coincides with aforementioned observation of the steady-state flow curves (Fig.1b), although a rigorous theoretical explanation demands further work. And such rate dependence appears insensitive to the exact value of $\phi$, as long as it is high enough to develop a solidity.

It is also worth noting that, once the volume fraction goes below the jamming point and the system becomes a true suspension, the evolution of images is dramatically different. This is demonstrated by Fig.3c, with further information on the solid-fluid transition to be discussed in the next section.

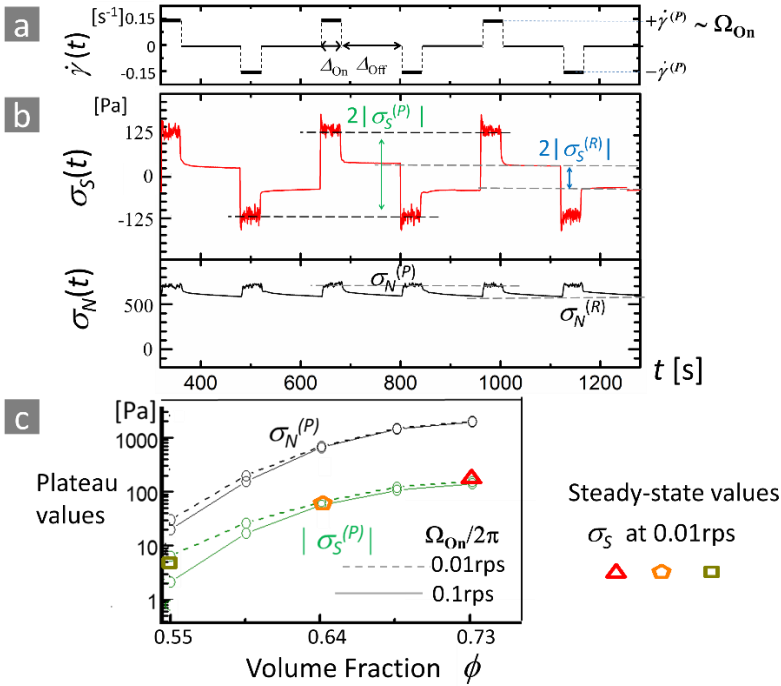

**Fig. 5** Stress measurements with cyclic shear cessations ---- **(a)** Protocol of CSC as demonstrated by the time-dependent shear rate $\dot{\gamma}(t)$; **(b)** The time-dependent shear and normal stress in response to the CSC, as well as illustrations of the plateau values, $|\sigma_S^{(P)}|$ and $\sigma_N^{(P)}$, and stress residues, $|\sigma_S^{(R)}|$ and $\sigma_N^{(R)}$. Sample signals are shown for the case of $\phi$ =0.73, with a smoothing by 0.1s to suppress the random noise. **(c)** Plateau values as functions of $\phi$, in experiments with two different values of $\dot{\gamma}^{(P)}$, $\Delta_{Off}$=120s, and accumulated strain $\dot{\gamma}\,\Delta_{On}$ = 5.5 for each half-cycle. Statistical variations among different cycles are all within the size of the symbols. Corresponding steady-state stress (adapted from Fig.1c) are also displayed for comparison.

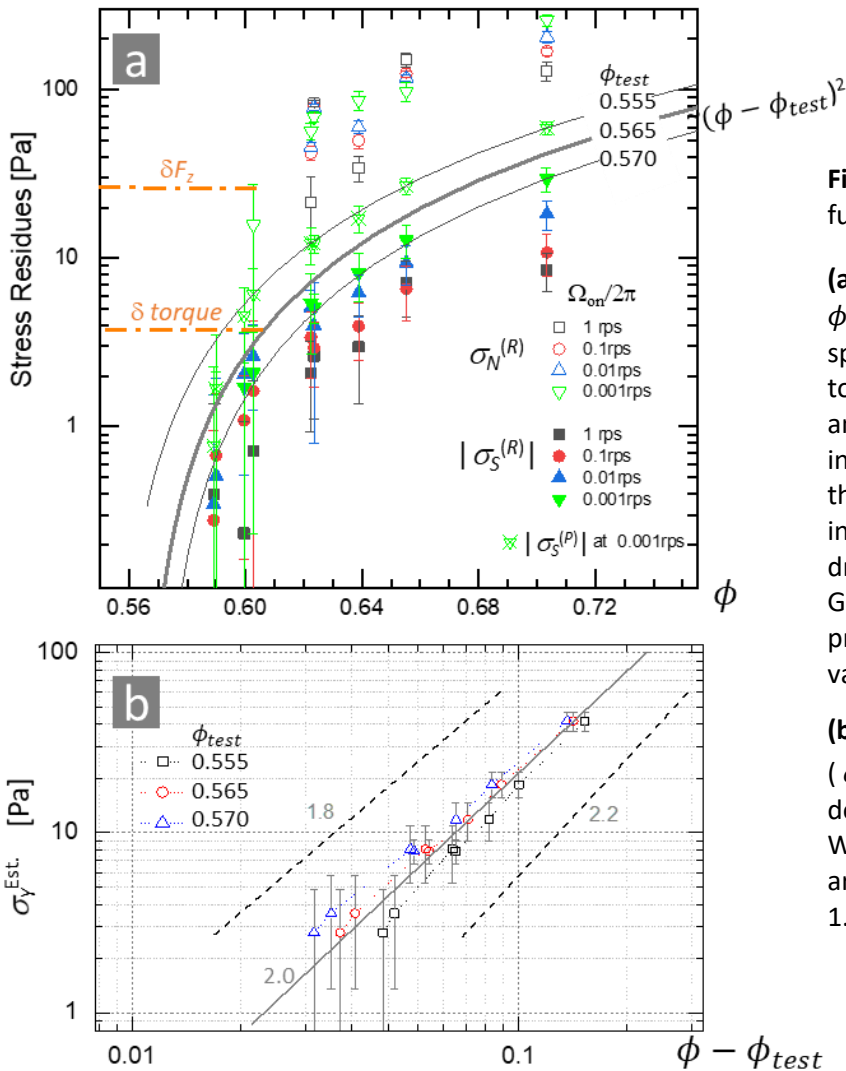

**Fig. 6** Stress residues and estimated yield stress, as functions of volume fraction

**(a)** Stress residues $|\sigma_S^{(R)}|$ and $\sigma_N^{(R)}$ as functions of $\phi$ in log-linear scales, for four different driving rates spanning over three decades. Uncertainties $\delta$ due to the different levels of noise in measuring *torque* and normal force (F$_z$), respectively, are also indicated by the dot-dash lines. Error bars represent the standard deviation among different cycles. Also included is the plateau stress $|\sigma_S^{(P)}|$ at the slowest driving rate 0.001rps, as an additional reference. Gray lines shows the hypothetical curves proportional to $(\phi - \phi_{test})^2$, for three different values of $\phi_{test}$ as indicated.

**(b)** Our best estimate of the yield stress $\sigma_Y^{Est.}$= $(\sigma_S^{(P)} * \sigma_S^{(R)})^{0.5}$, using the data of 0.001rps. Such definition corresponds to the thick gray line in (a). With three different values of $\phi_{test}$, the same data are plotted against $\phi - \phi_{test}$. Logarithmic slopes of 1.8 , 2.0, and 2.2 are also displayed as references.

# Solid-fluid transition over the change of volume fractions

## 1. Response to cyclic shear cessations (CSC)

To supplement the studies of steady states, we perform experiments with cyclic shear cessations. Illustrated in Fig.5a, our CSC is defined by two characteristic times, $\Delta_{On}$ and $\Delta_{Off}$, plus a constant rate $\dot{\gamma}^{(p)}$ that is set by the rotation rate $\Omega_{On}$ imposed in alternating directions. Fig.5b shows the typical response in the effective stress in analogy to that defined in steady-state experiments. The nearly periodic response provides two pairs of characteristic values. (1) In the mobile period, the system undergoes a short transient (in comparison to $\Delta_{On}$) and reach two plateaus ($|\sigma_S^{(P)}|$, $\sigma_N^{(P)}$). (2) The stress does not just vanish after the abrupt stops of the driving boundary. Rather, two residues ($|\sigma_S^{(R)}|$, $\sigma_N^{(R)}$) remain as a "memory" for the previous direction[22]: These non-zero residues show that our tight packing is an interesting contrast to an over-damped suspension (--- see Fig.1 of the cited reference), in which sudden cessations of shearing leave no residues but the memory shows up upon the starting of the next mobile period.

Figure 5c presents our measurements of plateau values as functions of $\phi$ at two rotation rates, for both $|\sigma_S^{(P)}|$ and $\sigma_N^{(P)}$. In addition, we adapt data of steady-state stress $\sigma_S$ from Fig.1 to show that, with an accumulated strain $\dot{\gamma} \, \Delta_{On}$ = 5.5 for each half-cycle, the values of $|\sigma_S^{(P)}|$ in our CSC are in fairly good agreement with the true steady-state stress.

In the attempts of approaching the solid-fluid transition, we perform series of CSC experiments and measure the stress residues with care, using four different rotation rates spanning over three decades, shown as in Fig.6. The rationale are the following: (A) For our system with hydrogels, we have verified that both $\sigma_S$ and $|\sigma_S^{(P)}|$ decreases monotonically as the shear rate is reduced, within the range of our observations. (B) It is shown in previous studies of microgels that the stress residue increases over the decrease of shear rate [3], and we have also verified that this is also true with our fluid-immersed macroscopic hydrogel particles --- see data of $|\sigma_S^{(R)}|$ in Fig.6a. (C) It is only natural that the dynamical yield stress $\sigma_Y$ must be bounded by the relationship $|\sigma_S^{(R)}| < \sigma_Y < |\sigma_S^{(P)}|$, and that $\sigma_Y$ can estimated simply by progressively reducing the driving rate (not necessarily to zero). Therefore, by measuring how $|\sigma_S^{(R)}|$ or $|\sigma_S^{(P)}|$ changes as functions of $\phi$, one shall be able to get a reasonable estimate of how $\sigma_Y$ approaches zero. In fact, as shown by Fig.6, we find that both $|\sigma_S^{(R)}|$ and $|\sigma_S^{(P)}|$ present a good fit to a be proportional to $(\phi - \phi_c)^2$. However, the uncertainty on $\phi_c$ is significant, because instrumental noise has limited reliable stress measurements only at values of $\phi$ that are at a substantial distance above the critical value. The conclusion is sensitive to the model and its extrapolation.

## 2. Direct evidence of solid-fluid transition, by imaging

On the contrary, information from imaging does not degrade along the path of approaching the solid-fluid transition as the stress measurements might do. In fact, using the previously described protocol, we observe the settling of particles in response to shear cessations with small incremental changes of $\phi$. In Fig.7, the data at three different driving speeds consistently show that a qualitative transition must have occurred between $\phi = 0.549$ and $0.554$: Above the transition, the long exponential tails feature the adjustment of particles with the relaxation of the elastic network. Below the transition, the system becomes a suspension. The settling of the neutral-buoyant particles is dominated by the dynamics of interstitial fluid, which features a rate-dependent transient state involving effects from both inertia and viscous damping. Video demonstrations are available online (http://www.phys.sinica.edu.tw/jctsai/relax/ ).

Intriguingly, the solid-fluid transition in our system shows a critical volume fraction that coincides with the random-loose-packing (RLP) value reported in prior studies [23]. However, as RLP is usually discussed in the context of frictional packing, whether there is a real connection between the solid-fluid transition of our system which is well lubricated, and the value of RLP, demands further studies to clarify.

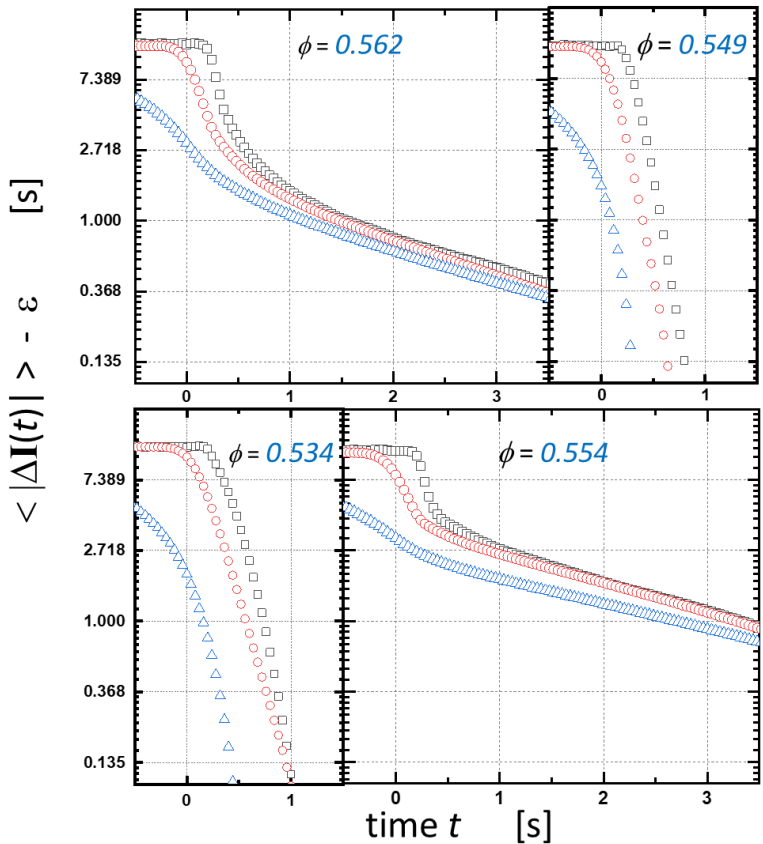

**Fig.7** Settling of particles, in response to the cessation of shearing at $t=0$, for four different volume fractions $\phi$ as indicated. Definitions of $<|\Delta\mathbf{I}(t)|> - \varepsilon$, and time $t$ follow those from Fig.3. Data include runs with $\Omega_0(t<0)/2\pi = 1$rps (squares), $0.1$rps (circles), and $0.01$rps (triangles), respectively.

## Discussion

### 1. Transition of Reynolds number, and its influence on determining the yield stress

Many prior experiments[1,12] have adopted Herschel-Bulkley model (HB, $\sigma_S = \sigma_Y + K\dot{\gamma}^n$) and successfully determined a well-defined jamming density by analyzing the transition of $\sigma_Y$. However, the HB-fit with our data presents certain difficulties, such as at low shear rates and for $\phi$ near the solid-fluid transition. One reason that HB model is unsuitable for our analyses involves the range of particle Reynolds number ($Re_p$). In principle, the solid-fluid transition of the change of volume fractions can be determined by the vanishing of the *dynamical yield stress* $\sigma_Y$, which in convention relies on measuring the flow curve $\sigma_S(\dot{\gamma})$ at small values of $\dot{\gamma}$ where the force mediated by the interstitial fluid is negligible in comparison to that from direct contacts. However, due to the low elastic modulus of our hydrogel particles, the stress contributed purely by direct contacts can go below the instrumental noise as $\phi$ gets close to the transition --- see Fig.6a. As a result, such determination of $\sigma_Y$ counts, in a large part, on the extrapolation of $\sigma_S(\dot{\gamma})$ from higher values of $\dot{\gamma}$ where hydrodynamics matters. But unlike other experiments performed in microfluidic setups [12] at low Reynolds numbers, our particles are centimeter-sized. To obtain significant force signals for $\sigma_S(\dot{\gamma})$, the particles Reynolds number can easily exceed unity, unless we opt to make the interstitial fluid extremely viscous. There have been no lacks of prior experiments and simulations [9] in contexts of high viscosity, which could make the yield stress $\sigma_Y$ undetectable evet at very small shear rates. Those studies present excellent cases for the commonly cited "J rheology"[10], as well as the "K rheology" with inertia included [11]. Both paradigms are based on the simple Newtonian relationship $\sigma_{S,N} \sim \dot{\gamma}^1$, which excludes a nonzero $\sigma_Y$ as $\dot{\gamma}$ approaches zero.

We choose a low viscosity for our interstitial fluid, which has made the effect of yield stress explicit. However, due to instrumentational limitations described above, an accurate determination of the dynamical yield stress from flow curves would count on extrapolations and complications arise: The convexity of these curves stand, in a large part, more for the weighting between inertia and viscosity changes as the driving rate goes across $Re_p=1$, than for the emergence of yield stress. To model the flow curves correctly, one might have to use a less restrictive HB model, in which the index $n$ also depends on the Reynolds number (or $\dot{\gamma}$) --- and that is beyond the scope of our discussion. In this work, we attempt two alternative routes, by observing (a) the measurement of residual stress and (b) the particle settling after the sudden cessation of driving. It turns out that the latter provides a more sensitive detection of the solid-fluid detection than the former.

### 2. Long-time weakening of the hydrogel – possible effects on the relaxation of a packing

To understand the long-time relaxation of packing, we have performed an experiment with a long stall added to the end of our typical CSC protocol (of 25 cycles). Figure 8a presents the evolution of the shear and normal stress, respectively, as well as the stress ratio as functions of time. Note the logarithmic scale in time. The shear stress decays faster in the first three decades than in the last three. On the other hand, the decay of normal stress has a different feature and appears to become steeper beyond $10^2$s. Both $\sigma_S$ and $\sigma_N$ have decreased by more than 25 percent between $10^2$s and $10^5$s.

The shear-to-normal stress ratio, $\sigma_S/\sigma_N$, shows a rapid initial decay in the first $10^2$s, but is stabilized beyond $10^3$s. The rapid decay implies a reduction of structural anisotropy due to particle rearrangements, which we will demonstrate with internal imaging subsequently. The substantial decay of total stress between $10^2$s and $10^5$s but with a fixed stress ratio, on the other hand, suggests the weakening of the hydrogel particles themselves with an isotropic decay of mean stress over time. In what follows, we test our conjecture on the weakening of hydrogel with an experiment at the single-particle level.

To probe whether the strength of individual grains gradually weakens, we monitor the response of a single hydrogel particle to a fixed compressional strain over long time. Shown as in Fig.8b, the particle is compressed to $d - \varepsilon_0$ and stay fixed. The initial condition $F(0+)=0.6$N is about ten times of $d^2\sigma_N$ to mimic the situation of those load-bearing particles inside a high-volume-fraction packing. Multiple long-time experiments confirm that (1) the hydrogel maintains its strength up to about 100s, and that (2) hydrogel particles do weaken significantly at larger timescales. The fact (2) confirms our conjecture that material weakening is involved in the long-time decay of stress. In addition, the fact (1) in combination of the stabilized stress ratio at $10^2$s justifies our choice of $\Delta_{Off}$ =120s (30s) for evaluating the residual stress in Fig.5 (Fig.6).

We anticipate that such mechanical weakening can also play a role in steady-state flows at extremely slow shear rates. This suggests that the conventional strategy of obtaining the dynamical yield stress by extending window of observation in $\dot{\gamma}$ toward the "slow limits" might involve undesired complications for materials like hydrogels, which by themselves evolve slowly.

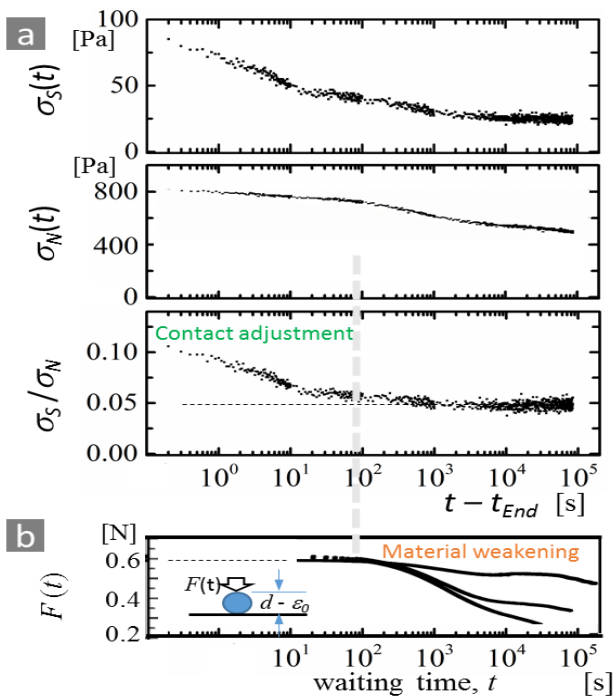

**Fig. 8 (a)** Time evolution of stress residues following a long stall, in terms of the shear stress, normal stress, and their ratio. The packing has a volume fraction 0.73. The offset $t_{End}$ marks the last stop of a CSC protocol of 25 cycles in which the magnitude of shear rate in the mobile periods is $0.14$s$^{-1}$. **(b)** Time evolution of the normal force $F(t)$ measured for a single particle at a fixed uni-axial compression $\varepsilon_0$. The particle is immersed in the same fluid as that in (a), stress-free for $t < 0$, and a compression is imposed to create an initial condition $F(0+)=0.6$N that stays constant at the timescale of 10-100s. The graph shows the data from three randomly chosen particles.

## Conclusion

In this work, we study the granular dynamics of soft, frictionless particles in response to shearing in a fixed volume. Particles are fully immersed in fluid with density and refraction-index matched. Use of sudden cessations of shearing proves to be good supplements to the conventional steady flow-curve measurements, in decoupling the fluid dynamics from effects of direct contacts between particles. In particular, internal imaging provides fruitful information in (a) the rate-dependent dynamics, reflecting a change of dominance from quasi-static yielding in the slow regime, to smooth fluid-mediated sliding in the fast regime; and (b) an accurate determination of the solid-fluid transitions over incremental changes of volume fraction $\phi$. In addition, we show that at a slow driving, a few characteristic values of stress such as the steady (plateau) stress ($\sigma_S^{(P)}$), the residues ($\sigma_S^{(R)}$), and the estimated yield stress ($\sigma_Y^{Est.}$) consistently follow a power law with an exponent very close to 2, in terms of the "distance to jamming". [12–14]

One intriguing question remains open: What determines the transitional zone of shear rates (between 0.1 and 1 s$^{-1}$) for the rate-dependent behaviors that consistently occur in both the conventional flow-curve measurements (Fig.1) and the response to sudden cessations (Fig.3-4)? Inverting the transitional shear rates gives the timescale between 10 and 1s. However, with the material parameters in our experiment, time constants constructed from dimensional analyses such as $\eta/\sigma_N$ or $\eta/E$ are well below 10$^{-5}$s. These values are obviously too small to account for the relaxation we have observed, even though they are commonly used as the time unit in theories and simulations[24]. We believe that a proper explanation on what we find must include the dynamics of lubrication and, most likely, in conjunction with *draining* between particle surfaces and/or the permeability of hydrogel. Full answers demand further works. However, there are a few clues: Our internal imaging of steady-state flows at different driving rates do reveal a non-smoothness of particle trajectories that depends on shear rate. We have also noted that such rate dependence can become dramatic when slippery particles are replaced with frictional grains [25]. Another interesting question is about the subtle rate dependence of the total relaxation ($f_0$), which we leave open in describing the rate-dependent settling of particles at the slow limit.

We hope that our survey of the shear flow and relaxation of frictionless soft particles provokes questions worthy of attention, and sheds lights on extending the existing paradigms for understanding the macroscopic rheology of *packed grains* by taking into account both the *finite rigidity* and fluid-mediated interactions between constituent particles.

-----------------------------------------------------------------------------------